\documentclass[manuscript]{acmart}

\AtBeginDocument{%
  \providecommand\BibTeX{{%
    \normalfont B\kern-0.5em{\scshape i\kern-0.25em b}\kern-0.8em\TeX}}}

\usepackage{booktabs}
\usepackage{multirow}
\usepackage{tikz}
\usetikzlibrary{shapes,arrows}
\usepackage{amsmath,bm,times}
\usepackage{verbatim}
\usepackage{adjustbox}
\usepackage{wrapfig}
\usepackage{graphicx}
\usepackage{subfig}
\newlength{\myMheight}
\copyrightyear{2020}
\acmYear{2020}
\setcopyright{acmcopyright}\acmConference[RecSys '20]{Fourteenth ACM Conference on Recommender Systems}{September 21--26, 2020}{Virtual Event, Brazil}
\acmBooktitle{Fourteenth ACM Conference on Recommender Systems (RecSys '20), September 21--26, 2020, Virtual Event, Brazil}
\acmPrice{15.00}
\acmDOI{10.1145/3383313.3412211}
\acmISBN{978-1-4503-7583-2/20/09}

\begin{document}
\title[Interpretable Contextual Team-aware Item Recommendation]{Interpretable Contextual Team-aware Item Recommendation: Application in Multiplayer Online Battle Arena Games}


\author{Andr\'es Villa}
\email{afvilla@uc.cl}
% \authornote{Corresponding authors}
\author{Vladimir Araujo}
\email{vgaraujo@uc.cl}
\orcid{0000-0001-5760-8410}
\additionalaffiliation{%
  \institution{Millennium Institute Foundational Research on Data, IMFD}
  \city{Santiago}
}
\author{Francisca Cattan}
\email{fpcattan@uc.cl}
\author{Denis Parra}
\email{dparra@ing.puc.cl}
\orcid{0000-0001-9878-8761}
\authornotemark[1]
\affiliation{%
  \institution{Pontificia Universidad Cat\'olica de Chile}
  \city{Santiago}
}


\begin{abstract}
The video game industry has adopted recommendation systems to boost users interest with a focus on game sales. Other exciting applications within video games are those that help the player make decisions that would maximize their playing experience, which is a desirable feature in real-time strategy video games such as Multiplayer Online Battle Arena (MOBA) like as DotA and LoL. Among these tasks, the recommendation of items is challenging, given both the contextual nature of the game and how it exposes the dependence on the formation of each team. Existing works on this topic do not take advantage of all the available contextual match data and dismiss potentially valuable information. To address this problem we develop TTIR, a contextual recommender model derived from the Transformer neural architecture that suggests a set of items to every team member, based on the contexts of teams and roles that describe the match. TTIR outperforms several approaches and provides interpretable recommendations through visualization of attention weights. Our evaluation indicates that both the Transformer architecture and the contextual information are essential to get the best results for this item recommendation task. Furthermore, a preliminary user survey indicates the usefulness of attention weights for explaining recommendations as well as ideas for future work. The code and dataset are available at: {\textit{\href{https://github.com/ojedaf/IC-TIR-Lol}{https://github.com/ojedaf/IC-TIR-Lol}}}.
\end{abstract}




\maketitle

\section{Introduction}
The annual report of Newzoo shows that global e-sports revenues and its audience will grow to \$1.1 billion and 495 million people in 2020, respectively \cite{report}. MOBA is one of the most significant social gaming genres contributing to that growth. An example of this phenomenon is the League of Legends World Championship, which was the biggest tournament of 2019 with more than 105 million hours live on Twitch and YouTube. This type of games can rise up to 64 million active players per month worldwide, with over a billion monthly gaming hours \cite{players}. Most of its popularity is due to social dynamics that motivate new players to engage in long-term commitment to the game \cite{Tyack:2016:AMG:2967934.2968098}.

In this context, several studies have leveraged artificial intelligence to recommend videogames \cite{cheuque2019recommender}, as well as to improve the personal experience of players, in applications like difficulty adjustment \cite{Silva2017,Araujo2018}, intelligent agents \cite{openai2019dota}, and in-game recommender systems \cite{Araujo:2019:DMI:3298689.3346986,Chen2018}. Regarding recommender systems, one challenge is to suggest the users the most suitable set of items for their characters considering the context of a specific match. Existing approaches attempt to solve the problem simply by using character descriptors, thus ignoring relevant contextual information from matches. Also, they focus on recommending a single character. However, such recommendations do have a common goal and a group (team) recommendation may be appropriate.

In this paper, we focus on exploiting  contextual information present in each match in order to generate richer representations of the characters, thus improving item recommendations for each participant in a team. Such information corresponds to the specific champion used, the role, and the team that each player belongs to. Inspired by the Transformer neural architecture \cite{vaswani}, we propose to use its encoder layer to model the relationships between the descriptor vectors for each of the aforementioned features. Additionally, its multi-headed attention mechanism helps to acquire information that makes it possible to interpret what the model is focusing on. We extensively evaluate our system by conducting comparisons with state-of-the-art methods on a real and challenging dataset. We also conduct a preliminary user survey to gain insights about the recommendation performance and the usefulness of attention-based explanations.

The contributions of our work are: (i) Introducing the method TTIR (Team-aware Transformer-based Item Recommendation), which significantly outperforms existing works on several ranking metrics and provides support to the importance of the team and role contexts, (ii) Designing a visual explanation mechanism in order to help users understand and follow team-aware item recommendations, and (iii) Providing ideas for future work by conducting a preliminary user survey to gain insights from the quality of the recommendation and the explanations provided.

\section{MOBA Games: Overview and Recommendation Problem}\label{Problem}

\begin{wrapfigure}{R}{0.35\textwidth}
\centering
\vspace{-4mm}
\includegraphics[width=0.3\textwidth]{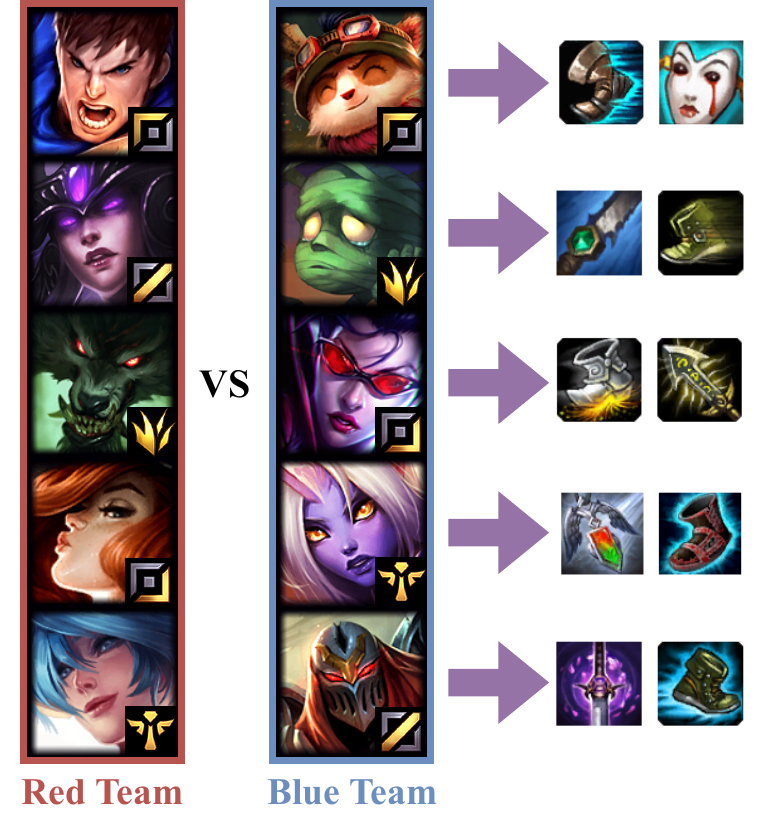}
\vspace{-3mm}
\caption{\label{fig:problem} Example of two teams matchup and item recommendations for the Bue team. Symbols on the bottom right corner of each champion represents their role.}
\vspace{-6mm}
\end{wrapfigure}

The MOBA genre corresponds to strategy video games in which each player controls a single character as part of a team competing against another team of players on a battle arena. Among the different video games cataloged under the MOBA genre, League of Legends (LoL) has dominated the market since 2012 and is considered one of the most popular electronic games worldwide. Each match consists of ten players divided into two teams of five people (\textit{Blue} and \textit{Red}). The main goal of the game is to battle head-to-head across a fixed battlefield to destroy the base of the enemy team. Each player selects their unique champion from a total of more than 146 available, according to the player's preferences and the composition of the allied team.

The pace of the game is encouraged by an in-game currency reward system, which is used to buy items that increase the statistics and performance of the champion. This is one of the main ways for the players to increase their attack and defense power, thereby increasing their contribution to winning the game. 
Players can choose up to six items from approximately 233 available. However, several of the items can be combined to obtain a total of stronger 89 finished items, which we use in this work.
Both the choice of champions and the items pose the challenge of the number of possible combinations, which users face making decisions based on experience.
This is particularly complex for new players and presents interesting opportunities for in-game recommender systems.

\section{Related Work}\label{Related}
\textbf{In-game recommendation for MOBA games}. In recent years, methods for in-game recommendations have received interest, where most works focused on character suggestion \cite{Chen2018,8711985,8986550}. However, there has been little work on item recommendation, recently showing two approaches based on data mining methods. One for the recommendation of the future item given an initial set of items \cite{8395021} and another for the recommendation of a fixed item set \cite{Araujo:2019:DMI:3298689.3346986}. We closely follow the methodology from \cite{Araujo:2019:DMI:3298689.3346986}; however, unlike their approach that uses only a few attributes of the data, we leverage meaningful contextual information about the game such as the allies, enemies, and the role of champions.

\textbf{Group recommendation}. The increase in social networking increased the importance of group recommendations in various domains \cite{masthoff2011group,Amatriain2015}. The most common systems are applied to recommendation of movies \cite{7226695}, music \cite{10.1016/j.eswa.2014.11.042}, and travel \cite{6521356}. All those systems attempt to recommend products or services to a group that has a common aim at a particular moment while increasing the individual satisfaction of each user. Another interesting domain is videogames, which is still an open issue. Recently, a multi-profile team-based recommender system for PvP games was proposed \cite{10.1145/3290688.3290750} to help teams improve by suggesting play styles and weapons to use. That approach is not directly related to our proposal because it was applied for a MMOG game using user profile data. Instead, we focus on the MOBA genre with an approach that does not use information from the user but from the characters in the game for item recommendations.

\textbf{Recommendation systems with Transformer}. Recently, the Transformer \cite{vaswani} has been a foundation for many competitive methods. This architecture has been shown to efficiently encode various types of information useful for recommendation systems \cite{Zhou_2020}. A personalized re-ranking model is proposed in \cite{Pei2019}, which captures in its encoding layer the interactions between users and items to produce an interpretable re-ranked recommendation list. Other works use this model, including the user's behavior sequence to learn more in-depth representations for each item in the sequence \cite{chen2019behavior, Chen:2019}. Unlike previous works we apply it to in-game interpretable item recommendation with newer contexts. 

\section{Model Architecture}

\begin{figure}[h]
\vspace{-3mm}
\includegraphics[width=0.85\linewidth]{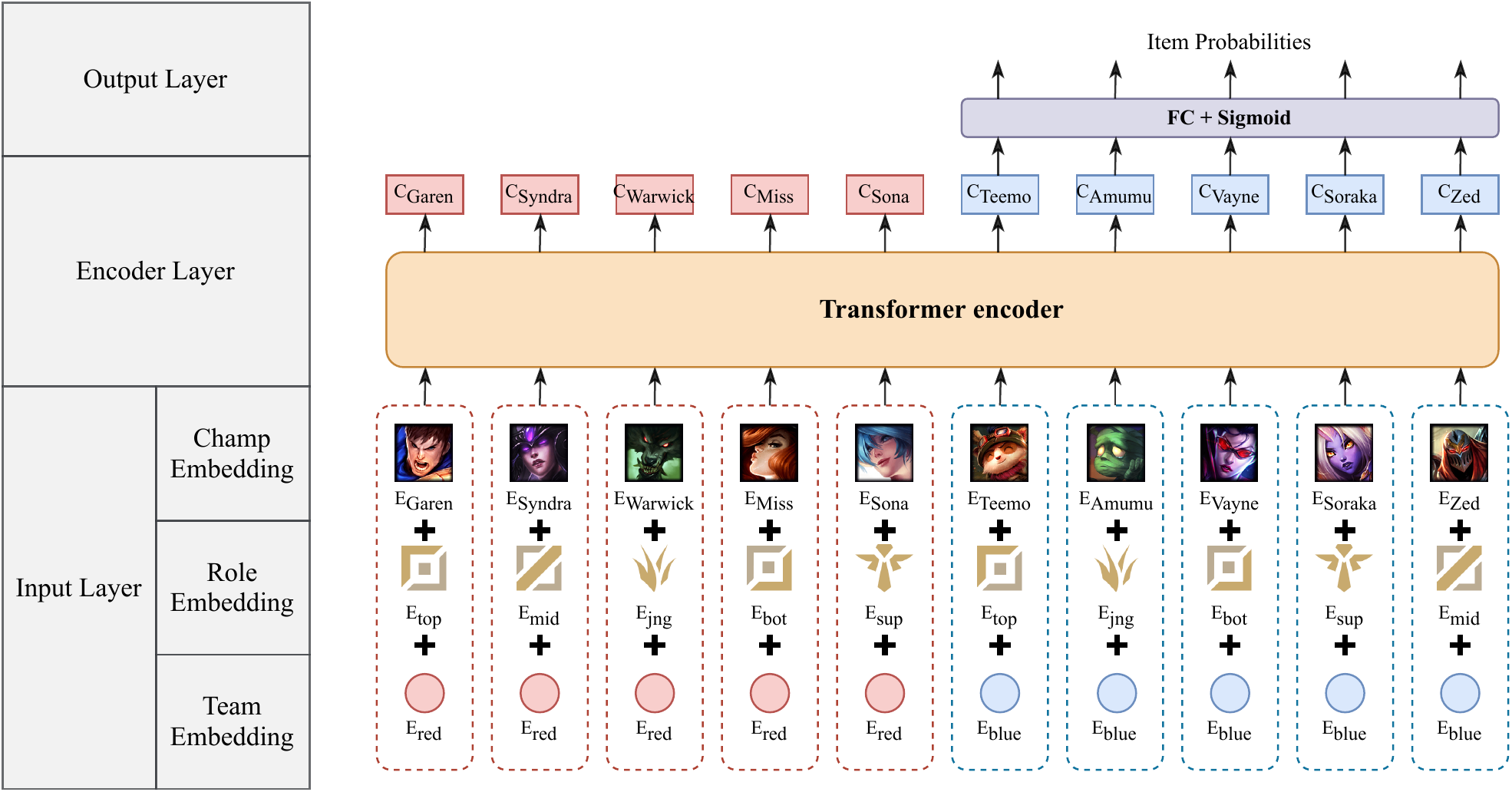}
\vspace{-3mm}
\caption{Network architecture of TTIR.}
\label{fig:model}
\vspace{-3mm}
\end{figure}

Figure~\ref{fig:model} shows the \textbf{T}ransformer for \textbf{T}eam-aware \textbf{I}tem \textbf{R}ecommendation architecture (TTIR). This model is made up of three major parts: the input representation layer, the encoder layer, and the output layer for recommendation. It takes as input the information of a match, which consists of the champions, their assigned role, and the team they belong to. Then it recommends a list of six items to each of them. The details of our architecture are in the following paragraphs.

\textbf{Input Layer}. The goal of the input representation layer is to prepare an unambiguous representation for each participant within the match, considering its champion, role and the team they belong to. To represent these features we take inspiration from the BERT language model \cite{devlin-etal-2019-bert}. It represents sequences of words as sentences, adding to the representation of each word the information about its position and the sentence it belongs to. In the same spirit, we use three learned embeddings: the champion embedding $E_{champ}$, the role embedding $E_{role}$ and the team embedding $E_{team}$. At last we add all of these embeddings to obtain the champion embedding $E_{input}$, which has the model dimension $d_{model}$ as shown in the equation~\ref{equ:input_repre}. The lower part of the Figure~\ref{fig:model} shows a visual example of the input representation. 

\vspace{-2mm}
\begin{equation}
\label{equ:input_repre}
E_{input} = E_{champ} + E_{team} + E_{role}, \;where\;E \in \mathbb{R}^{d_{model}}
\end{equation}

\textbf{Encoder Layer}. This part of the model is a Transformer encoder based on the original implementation \citep{vaswani}. We omit an exhaustive description of the Transformer architecture because its use has become ubiquitous in recent years. The goal of this encoder is to compute interactions between allied and enemy champions of the match with the self-attention mechanism. The output of this encoder are contextualized embeddings $C$ of the input embeddings $E_{input}$ that captures complex relations between the champions. This architecture has two principal hyper-parameters, the number of layers $l$ and number of heads $h$. The influence of these parameters is studied later in the ablation analysis.

\vspace{-2mm}
\begin{equation}
\label{equ:att}
C = TransformerEncoder(E_{input}), \;where\;C \in \mathbb{R}^{d_{model}}
\end{equation}

\textbf{Output Layer (Item Recommendation)}. The purpose of the item recommendation layer is to generate a list of items for each champion of a team in a match. In order to do this, we feed each contextualized embedding \textit{$C^{i}$} of a team, where \textit{i} denotes the champion, through a linear layer, followed by a sigmoid function. As it is shown in the equation~\ref{equ:prob}, the final output are the probabilities that the champion selects each of the items in $N_{items}$. Given that the champions can only use up to six items, the model recommends the six most probable items for each of the five players in a team (equation~\ref{equ:item_rec}). 

\vspace{-2mm}
\begin{equation}
\label{equ:prob}
P_{items}^{i} = Sigmoid\left(W_{rec} C^{i}\right), \; {W}_{rec}  \in R^{  N_{items} \times d_{model}}, C^i \in R^{d_{model}}, i \in [1,5]
\end{equation}

\vspace{-3mm}
\begin{equation}
\label{equ:item_rec}
Recommendation^{i} = top\left(P_{items}^{i}, 6\right)
\end{equation}

Like \citep{Araujo:2019:DMI:3298689.3346986}, the model was supervised only with the items selected by each champion of the winning team. This way the likelihood of recommending the best items is maximised in order to win. 

\section{Experiments}\label{Experiments}
In this section, we describe the dataset used and the offline evaluation conducted on the final TTIR model. Along with the results, an ablation study is presented to better understand the behavior of the network.

\textbf{Dataset}. In order to train and evaluate our model, we used a publicly available dataset provided by Kaggle\footnote{www.kaggle.com/paololol/league-of-legends-ranked-matches}. This consists of 184,070 game sessions in the ranked category, a competitive alternative to the normal match. Although the dataset does not provide a specific structure for recommendation tasks, we adequate it for this purpose. The raw dataset includes several files with much information about each match, so we choose the most relevant for this work. Specifically, our final dataset contains a match per instance, consisting of the identifier of each of the 10 participants with their role, team, and items used. We filter the basic, advanced, and consumable items, as well as all the matches that did not belong to the LoL 7th season. The complete data was divided into two subsets for training and testing, taking into consideration that the least common champion must be present in both of them. 
The overview of the final dataset is shown in Table \ref{tab:dataset}.

\begin{table}[]
\centering
\small
\caption{Overview of the dataset}
\label{tab:dataset}
\vspace{-3mm}
\def\arraystretch{0.9}
\begin{tabular}{@{}lc@{}}
\toprule
 & \textbf{LoL Ranked Matches 7th Season} \\ \midrule
\# Items & 89 \\
\# Champions & 136 \\
\# Matches & 157,584 \\
\ Roles & Top \includegraphics[height=\myMheight]{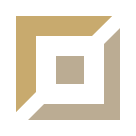}, Mid \includegraphics[height=\myMheight]{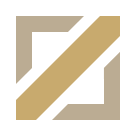}, Jungle \includegraphics[height=\myMheight]{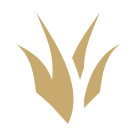}, Support \includegraphics[height=\myMheight]{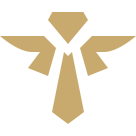}, Bot \includegraphics[height=\myMheight]{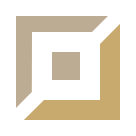} \\
\ Train / Test & 1,261,280 / 314,560 \\ \bottomrule
\end{tabular}
\vspace{-3mm}
\end{table}

\textbf{Training and Evaluation Settings}. The best configuration of TTIR consists on 2 heads, 1 layer, an embedding size of 512 and dropout of 0,5. We trained the model using Adam optimizer with a learning rate of 3e-4 until convergence. For evaluation, we compare our model with decision tree (D-Tree), logistic regression (Logit) and shallow artificial neural network (ANN) baselines of \cite{Araujo:2019:DMI:3298689.3346986}. We also implemented an additional baseline based on Convolutional neural networks (CNN) for a stronger comparison. We used different evaluation metrics to measure the relevance and ranking of our recommender system: Precision@k, Recall@k, F1@k, and MAP@k, with k=1,6,10. While k=6 might seem a strange cut, it is the maximum number of items a player can use during a match. Likewise, k=10 shows how the six articles are distributed in the firsts ten positions of the list.

\subsection{Results}

Results are shown in Table~\ref{tab:evaluation-results}. Our model achieves consistent and statistically significant improvements compared to the best baseline CNN. The trend indicates that the biggest performance difference is observed in top rank positions (k=1,6) and this difference decreases as the cut-off increases (k=10). This result is positive for TTIR, since it is difficult for a model to know when it has to recommend at least six items, as it normally depends on external factors like the game duration and the player expertise. The difference in performance between TTIR and other models is due to the ability of the Transformer to include relevant contextual information of the match into the representation of each champion.

\begin{table}[htbp]
  \centering
  \vspace{-2mm}
  \caption{Experiment Results}
  \vspace{-2mm}
  \resizebox{8cm}{!}{
  \subfloat[Results for top @k recommendation. TTIR is significantly better than the second best method, CNN.]{%
    \label{tab:evaluation-results}
    \setlength\tabcolsep{1.5pt} % default value: 6p
    \def\arraystretch{0.9}
    \begin{tabular}{@{}lcccccc@{}}
    \toprule
     &  \multicolumn{5}{c}{\textbf{Method}} & \textbf{T-test($df=314557$)}\\\cmidrule(l){2-6} 
    \multicolumn{1}{l}{} & \multicolumn{1}{l}{D-Tree} & \multicolumn{1}{l}{Logit} & \multicolumn{1}{l}{ANN} & \multicolumn{1}{l}{CNN} & TTIR & p-Value (t-Stat) \\ \midrule
    Precision@1 & 0.516 & 0.672 & 0.771 & 0.790 & \textbf{0.803} & 5.30e-20 (9.158)\\
    Recall@1 &  0.135 & 0.178 & 0.205 & 0.209 & \textbf{0.214} & 9.54e-18 (8.580)\\
    F1@1 & 0.210 & 0.277 & 0.318 & 0.331 & \textbf{0.338} & 7.23e-20 (9.124)\\
    MAP@1 & 0.516 & 0.672 & 0.771 & 0.790 & \textbf{0.803} & 5.30e-20 (9.158)\\
    % \bottomrule
    % Precision@3 & 0.466 & 0.540 & 0.650 & 0.660 & \textbf{0.674} & 1e-15 (23.226)\\
    % Recall@3 & 0.362 & 0.423 & 0.510 & 0.518 & \textbf{0.530} & 1e-15 (23.073)\\
    % F1@3  & 0.399 & 0.465 & 0.560 &  0.580 &\textbf{0.593} & 1e-15 (24.373)\\
    % MAP@3  & 0.654 & 0.750 & 0.831 & 0.844 &\textbf{0.854} & 1e-15 (23.998)\\
    \bottomrule
    Precision@6 & 0.319 & 0.393 & 0.476 & 0.484 &\textbf{0.492} & 2.41e-22 (9.723)\\
    Recall@6  & 0.491 & 0.607 & 0.732 & 0.744 &\textbf{0.756} & 1.93e-27 (10.854)\\
    F1@6 & 0.379 & 0.468 & 0.566 & 0.586 &\textbf{0.596} & 2.57e-27 (10.828)\\
    MAP@6 & 0.648 & 0.714 & 0.785 & 0.795 & \textbf{0.805} & 3.77e-30 (11.410)\\
    \bottomrule
    Precision@10 & 0.204 & 0.285 & 0.341 & 0.348 &\textbf{0.351} & 2.49e-11 (6.674) \\
    Recall@10  & 0.520 & 0.726 & 0.864 & 0.882 &\textbf{0.889} & 1.43e-24 (10.232) \\
    F1@10 & 0.289 & 0.403 & 0.481 & 0.499 &\textbf{0.503} & 3.34e-15 (7.878) \\
    MAP@10 & 0.636 & 0.672 & 0.743 & 0.754 & \textbf{0.764} & 1.32e-34 (12.270) \\
    \bottomrule
    \end{tabular}
  }}\hspace{0.2cm}
  \resizebox{6.5cm}{!}{
  \subfloat[Ablation study of TTIR]{%
    \hspace{.7cm}%
    \label{tab:ablation}
    \setlength\tabcolsep{1.5pt} % default value: 6p
    \begin{tabular}{@{}lcccc@{}}
    \toprule
     & \textbf{P@6} & \textbf{R@6} & \textbf{MAP@6}\\ \midrule
    Default ($h$=2, $l$=1)  & 0.492 & 0.756 & 0.805\\ \midrule
    Multiheads ($h$=1)  & 0.462  & 0.726 & 0.778\\
    Multiheads ($h$=4)  & 0.492  & 0.756 & 0.806\\
    Layers ($l$=2)  & 0.493 & 0.757 & 0.806\\
    Layers ($l$=3)  & 0.493 & 0.758 & 0.807\\
    %Static $E_{team}$  & 0.492 & 0.854 & 0.805\\
    \midrule
    w/o $E_{enemies}$ & 0.487 & 0.749 & 0.798 \\
    w/o $E_{role}$ & 0.484 & 0.742 & 0.794 \\
    w/o $E_{enemies}$, $E_{role}$ & 0.479 & 0.736 & 0.787 \\
    \midrule
    CNN w/ $E_{enemies}$, $E_{role}$ & 0.484 & 0.744 & 0.795 \\
    \bottomrule
    \end{tabular}
    \hspace{.3cm}%
  }}
\vspace{-4mm}  
\end{table}

\textbf{Ablation analysis}. To understand the influence of contextual dimensions as well as several hyperparameters of the Transformer model, we conducted an ablation analysis which results are presented in Table \ref{tab:ablation}. The performance of TTIR does not increase with bigger numbers of attention heads (by default $h=2$), but it declines when the heads decrease to $h=1$. This confirms the importance of focusing on the different features of each champion. Increasing the number of layers (by default $l=1$) to $l=2$ or $l=3$ has an almost negligible improvement, but with a bigger cost on the number of parameters. This suggests preserving the default number of layers. In terms of contextual dimensions, we notice that removing the $E_{role}$ context has a slightly higher impact than the role of the $E_{enemies}$ context, but removing both contexts makes the model perform even worse than CNN with these contexts. These results indicate that it is not only the Transformer architecture of TTIR which makes a difference in performance, but the combined effect of both architecture and contextual information what makes TTIR work.

\section{Preliminary User Survey}\label{Preliminary User Survey}

In order to get insights from LoL players about the relevance of our recommendations and the usefulness of the attention weights from TTIR to explain the suggestions, we designed a visualization to explain the recommendations. Based on this, we conducted a preliminary survey.

\begin{figure}[]
\includegraphics[width=0.9\linewidth]{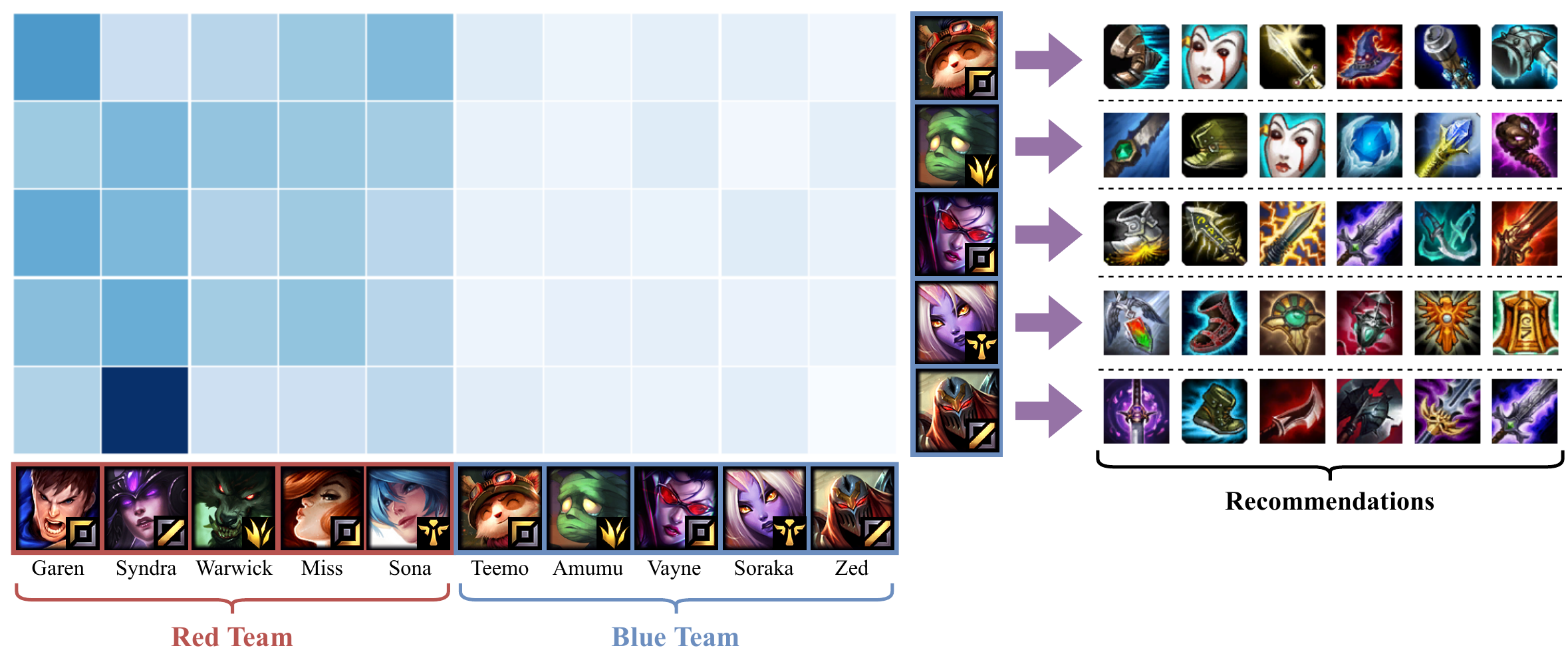}
\vspace{-4mm}
\caption{Visualization of the attention weights for each member of the Blue team on each member of both teams (bottom row).}
\label{fig:att}
\vspace{-3mm}
\end{figure}

\textbf{The visual explanation of team-aware item recommendations}. Figure \ref{fig:att} shows an example of attention weights visualization to explain the model recommendations. It consists of: (i) two teams, of five players each, at the bottom, (ii) a heat map in the center, and (iii) one of the teams with its six recommended items by each player on the right side. The heat map uses different color intensities to show the relevance of each champion upon each recommendation list. Darker colors represent more relevance. In Figure \ref{fig:att}, the model pays more attention to the enemies since the items are used to beat them, and the model recommends by maximizing the chance of winning.

\textbf{The survey procedure}. The survey consisted of showing subjects four cases similar to the one displayed in Figure \ref{fig:att}. Subjects were told that they belonged to the \textit{Blue team} and that they were facing the \textit{Red team}. Then they had to judge the quality of the recommendations for the \textit{Blue team} and the usefulness of the explanations provided by the heatmap. The survey was advertised in public Facebook group pages of college students who were LoL players. They replied their interest with an e-mail and we sent them back an online form where users had to agree with an informed consent about the study, and then answer the questions in Table~\ref{tab:results-survey} in a scale of 1 to 10 (1: completely disagree, 10: completely agree) for each of four cases similar to the one in Figure~\ref{fig:att}. We also asked open questions in order to get a less structured feedback from the participants: (Q4) \textit{If you \textbf{do not find} this visualization useful to explain the recommendation, tell us why}, and (Q5) \textit{If you \textbf{find} this visualization useful to explain the recommendation, tell us why}.

\textbf{Results of the user survey}. 16 people answered our survey. 25\% identified as female, 68\% as male and one subject did not disclose the gender. 13 out of 16 participants were between 20 and 30 years of age, one was between 18-20, the rest were older or did not disclose their age. Six subjects indicated that they started playing in the last 2-4 years, while the other 10 indicated having been playing for 6 years or more. Although we acknowledge having a small number of subjects (N=16), each one responded 4 cases of recommendations and their open comments provided evidence of paying deep attention to the user study. The results to questions Q1-Q3 are presented in Table \ref{tab:results-survey}. With respect to \textbf{Q1}, we observe that people have a fairly positive perception of recommendation relevance (M=7.98$\pm$1.22), but this perception seems to be more positive to less experienced users who started playing not before 2-5 years ago (M=8.46$\pm$1.3). In terms of \textbf{Q2}, we observe also a positive (M=7.4$\pm$1.42) and rather uniform impression among newer and experienced players, with respect to the perception of subjects towards influence of enemies and allies champions towards recommendation for the \textit{Blue team}. Finally, in \textbf{Q3} we notice that the perception of recommendation interpretability from the visualization is not as good as the perception of relevance (M=6.9$\pm$2.15), but again we observe a more positive impression from newer subjects (M=7.33$\pm$2.87) compared to the most experienced users. These results are consistent with previous studies indicating user expertise as a factor influencing perception about recommendations \cite{knijnenburg2011each,parra2015user}. To dig deeper into these results we analyzed the open user comments.

\begin{table}[h]
\vspace{-3mm}
\small
\caption{Results of the preliminary user survey (N=16), ratings in range [1-10]}
\vspace{-3mm}
\label{tab:results-survey}
%\begin{tabular}{l|cccc}
\begin{tabular}{@{}p{3in}cccc@{}}
\toprule
                                & & \multicolumn{3}{c}{Subjects by year of first play} \\ \cmidrule(l){3-5}
               & \textbf{Global M$\pm$SD} & \textbf{2009-11} & \textbf{2012-14} & \textbf{2015-2017} \\
\textbf{Question}          & (N=16) & (N=5) & (N=5) & (N=6)
\\ \midrule
Q1. How good were the recommendations for the \textit{Blue team} ?                              &  7.98$\pm$1.22 &  7.7$\pm$1.24 &  7.7$\pm$1.16 &  8.46$\pm$1.3                          \\ \hline
Q2. Is it understandable the influence of every team member upon each champion being recommended ?     &  7.44$\pm$1.72  &  7.4$\pm$1.55  &  7.1$\pm$0.8  &  7.75$\pm$2.49                     \\ \hline
Q3. Is it useful the information provided by the visualization in order to understand the item recommendations made ? &  6.9$\pm$2.15 &  6.7$\pm$1.98  &  6.6$\pm$1.65  &  7.33$\pm$2.87                          \\ 
\bottomrule
\end{tabular}
\vspace{-3mm}
\end{table}

\textbf{Synthesis of user comments}. On the positive side, we received comments of the usefulness of the explanations since they made sense to users based on their game experience: ``\textit{useful build to prevent enemy ganking...}'', ``\textit{you can see exactly the focus of each champion with respect to the main enemies on the facing team...}'', ``\textit{...with this build, Vlad hinders the enemy, making Lucian suffer. Then Ezreal with that build can damage both Lux and Fizz}''. We also received critical comments which provide important ideas for future work, several of them require information not currently available in our dataset: ``\textit{this explanation missed armor penetration and grievous wounds...}'', 
``\textit{it doesn't show with which item I need to start and the sequence to progress...}'', 
``\textit{recommendations do not show magic resistance...}'', 
``\textit{...in some cases, such as Soraka and Teemo, it would make more sense to show attention on the relationship with themselves}''.

\section{Conclusions}\label{Conclusions}

In this work we introduced TTIR, a contextual recommendation model which provides team-aware item recommendations in MOBA games such as LoL. TTIR successfully models the complex contextual relationships present in the matches, and the attention weights allow us to provide explanations of suggested items. Furthermore, with a preliminary user study we had an initial idea of the perception of relevance of actual LoL players as well as important feedback towards the visual recommendations based on TTIR attention weights. Our initial analysis indicates that expert users require more details for understanding and following the recommendations, while less experienced users find them coherent and useful. 
Among ideas for future work we consider providing further details in our recommendation explanations such as item statistics, as well as sequential item recommendation.

\begin{acks}
This work has been supported by the Millennium Institute for Foundational Research on Data (IMFD) and by the Chilean research agency ANID, FONDECYT grant 1191791.
\end{acks}

\bibliographystyle{ACM-Reference-Format}
\bibliography{references-moba}

\end{document}